\documentstyle[twoside,fleqn,espcrc2,epsfig]{article}

\newcommand{\AmS}{{\protect\the\textfont2
  A\kern-.1667em\lower.5ex\hbox{M}\kern-.125emS}}

\def\rmO{{\rm O}}

\def\proof{\noindent{\sl Proof:}\kern0.6em}

\def\frac#1#2{\hbox{$#1\over#2$}}
\def\dual{\mathstrut^*\kern-0.1em}

\def\lvec#1{\setbox0=\hbox{$#1$}
    \setbox1=\hbox{$\scriptstyle\leftarrow$}
    #1\kern-\wd0\smash{
    \raise\ht0\hbox{$\raise1pt\hbox{$\scriptstyle\leftarrow$}$}}
    \kern-\wd1\kern\wd0}
\def\rvec#1{\setbox0=\hbox{$#1$}
    \setbox1=\hbox{$\scriptstyle\rightarrow$}
    #1\kern-\wd0\smash{
    \raise\ht0\hbox{$\raise1pt\hbox{$\scriptstyle\rightarrow$}$}}
    \kern-\wd1\kern\wd0}

\def\nabstar#1{\nabla\kern-0.5pt\smash{\raise 4.5pt\hbox{$\ast$}}
               \kern-4.5pt_{#1}}

\def\drvstar#1{\partial\kern-0.5pt\smash{\raise 4.5pt\hbox{$\ast$}}
               \kern-5.0pt_{#1}}

\def\rhoprime{\rho\kern1pt'}
\def\rhobar{\bar{\rho}}
\def\rhobarprime{\rhobar\kern1pt'}
\def\rhobartilde{\kern2pt\tilde{\kern-2pt\rhobar}}
\def\rhobartildeprime{\kern2pt\tilde{\kern-2pt\rhobar}\kern1pt'}

\def\zetabar{\bar{\zeta}}
\def\zetaprime{\zeta\kern1pt'}
\def\zetabarprime{\zetabar\kern1pt'}
\def\zetar{\zeta_{\raise-1pt\hbox{\sixrm R}}}
\def\zetabarr{\zetabar_{\raise-1pt\hbox{\sixrm R}}}

\def\phiimpr{\phi_{\kern0.5pt\hbox{\sixrm I}}}

\def\ar{A_{\mbox{\scriptsize{\rm R}}}}
\def\vr{V_{\mbox{\scriptsize{\rm R}}}}

\def\aimpr{A_{\mbox{\scriptsize{\rm I}}}}
\def\vimpr{V_{\mbox{\scriptsize{\rm I}}}}

\def\diracstar#1#2{
    \setbox0=\hbox{$\gamma$}\setbox1=\hbox{$\gamma_{#1}$}
    \gamma_{#1}\kern-\wd1\kern\wd0
    \smash{\raise4.5pt\hbox{$\scriptstyle#2$}}}

\def\ba{b_{\rm A}}
\def\bv{b_{\rm V}}

\def\ca{c_{\rm A}}
\def\cv{c_{\rm V}}

\def\f1{f_1}
\def\kv{k_{\rm V}}
\def\kt{k_{\rm T}}

\def\kaaI{k_{\rm AA}^{\rm I}}

\def\CF{C_{\rm F}}

\def\opprime#1{\setbox0=\hbox{${\cal O}$}\setbox1=\hbox{${\cal O}_{\rm #1}$}
    {\cal O}_{\rm #1}\kern-\wd1\kern\wd0
    \smash{\raise4.5pt\hbox{\kern1pt$\scriptstyle\prime$}}\kern1pt}

\def\ophatprime#1{\setbox0=\hbox{$\widehat{\cal O}$}
    \setbox1=\hbox{$\widehat{\cal O}_{\rm #1}$}
    \widehat{\cal O}_{\rm #1}\kern-\wd1\kern\wd0
    \smash{\raise4.5pt\hbox{\kern1pt$\scriptstyle\prime$}}\kern1pt}

\def\bopprime#1{\setbox0=\hbox{${\cal O}$}\setbox1=\hbox{${\cal O}_{\rm #1}$}
    {\cal L}_{\rm #1}\kern-\wd1\kern\wd0
    \smash{\raise4.5pt\hbox{\kern1pt$\scriptstyle\prime$}}\kern1pt}

\def\blagprime#1{\setbox0=\hbox{${\cal B}$}\setbox1=\hbox{${\cal B}_{#1}$}
    {\cal B}_{#1}\kern-\wd1\kern\wd0
    \smash{\raise5.2pt\hbox{\kern1pt$\scriptstyle\prime$}}\kern1pt}

\def\mq{m_{\rm q}}

\def\za{Z_{\rm A}}

\def\zv{Z_{\rm V}}

\def\msbar{{\rm \overline{MS\kern-0.05em}\kern0.05em}}

\title{
{\vspace{-3cm} \normalsize
\hfill \parbox{30mm}{~}\\
\hfill \parbox{30mm}{CERN 97-256}\\
\hfill \parbox{30mm}{DESY 97-173}}\\
Non-perturbative O$(a)$ improvement of the vector 
       current
      \thanks{
              Talk given by M. Guagnelli at the International
              Symposium on Lattice Field Theory, 21$-$27 Juli 1997,
              Edinburgh, Scotland}
      }  
\author{M. Guagnelli\address{CERN, Theory Division, 1211 
                             Geneva 23, Switzerland}
        and
        R. Sommer\address{DESY-IfH, Platanenallee 6, D-15738 Zeuthen, Germany}}
       
\begin{document}

\begin{abstract}
We discuss non--perturbative improvement of the vector 
current, using the Schr\"odinger Functional formalism. 
By considering a suitable Ward identity, we compute the
improvement coefficient which gives the O$(a)$ mixing of the
tensor current with the vector current.
\end{abstract}

\maketitle

\section{INTRODUCTION}

It is well known that in Wilson's lattice QCD discretization errors
receive contributions that are linear in the lattice spacing. 
The most obvious recipe to reduce
them --- namely to go to the continuum limit by performing numerical
simulations at smaller lattice spacings  --- represents still a hard 
task, even in the quenched approximation. 

Following Symanzik~\cite{sym}, 
lattice artifacts can be removed order by order in
$a$ by adding appropriate higher--dimensional operators to the action
and to the fields whose correlation functions are of interest. If one
restricts oneself by requiring improvement only for on--shell
quantities~\cite{on_shell}, 
such as particle masses and matrix elements between
physical states, the structure of the improved action for QCD and of the
improved currents is rather simple.
The general theory was developed in detail in
ref.~\cite{definitions}. In the following we will use the same
notation without further reference.

In the quenched approximation, the improved action was determined
non--perturbatively for couplings $0 \leq g_0 \leq 1$ \cite{non_pert_act}.

The renormalized and improved axial--vector and vector currents are
parametrized as
\begin{eqnarray}
(\ar)^a_\mu &=& \za(1+\ba a\mq)(\aimpr)^a_\mu, \nonumber \\
(\vr)^a_\mu &=& \zv(1+\bv a\mq)(\vimpr)^a_\mu, \nonumber 
\end{eqnarray}
with
\begin{eqnarray}
(\aimpr)^a_\mu &=& A^a_\mu + a\ca \tilde{\partial}_\mu P^a,  
\; \tilde{\partial}_\mu = \frac{1}{2}(\partial_\mu + \partial^*_\mu) ,\nonumber\\
(\vimpr)^a_\mu &=& V^a_\mu + a\cv \tilde{\partial}_\nu T^a_{\mu\nu} \, .
\nonumber 
\label{eq:defcur}
\end{eqnarray}
In the above expressions, the coefficients $c_{\rm X}$ and $b_{\rm X}$ are
functions of the bare coupling, and are known to 1--loop order of
perturbation theory\cite{one_loop,pert_cv}. Moreover, the normalizations
($\za$ and $\zv$) and the improvement coefficients $\ca$ and
$\bv$ have been determined non--perturbatively in the quenched
approximation for $g_0 \leq 1$~\cite{non_pert_act,non_pert_coeff}. As
a next step, we determine $\cv$, thus completing
the knowledge of the improved vector current. The computation
of other improvement coefficients is discussed in \cite{giulia}.

Knowledge of $\cv$ is, for
example, required for the computation
of vector meson decay constants. The relative
contribution of $a\tilde{\partial}_\nu T_{\mu\nu}$ to these decay
constants can be, at $g_0^2=1$, as large as $0.3\times
\cv$~\cite{decay}.
Although the perturbative estimate~\cite{pert_cv},
\begin{equation}
\cv = -0.01225(1) \times g_0^2 \CF + \rmO(g_0^4),\, \CF=4/3,
\end{equation}
suggests an effect of less than 1\% on the decay constants, 
our preliminary non--perturbative
results determine
$\cv$ to be
much larger in magnitude for $g_0^2\simeq 1$.

\section{THE STRATEGY}

Chiral Ward identities relate correlation functions of axial--vector
and vector currents. In the O$(a)$ improved theory and for zero quark
mass these identities can be written in a form which is valid up to
error terms that are quadratic in $a$. Since the O$(a)$--improved
axial--vector current as well as
$\zv$ and $\za$
are known, one can use a particular Ward identity
in which the only unknown is $\cv$. 

Our starting point is the Ward identity (in the continuum):
\begin{eqnarray}
\int_{\partial R} {\rm d}\sigma_\mu(x) \epsilon^{abc}\langle
A^a_\mu(x)A^b_\nu(y){\cal Q}^c\rangle - \quad \qquad   \label{eq:ward1}\\ 
 2m\int_{R}{\rm d}^4x \epsilon^{abc} \langle P^a(x)A^b_\nu(y){\cal Q}^c\rangle
= 2i \langle V^c_\nu(y){\cal Q}^c\rangle \, , 
\nonumber
\end{eqnarray}
where the space--time region $R$ with boundary $\partial R$ contains
the point $y$ and ${\cal Q}^c$ is a source located outside $R$.

We then impose Schr\"odinger Functional boundary conditions
\cite{schr}, choose $\nu = k$ and specify the source,
\begin{equation} 
{\cal Q}_k^c = a^6\sum_{\bf y,z} \zetabar({\bf
  y})\gamma_k\frac{\tau^c}{2}\zeta({\bf z}) \, ,
\label{eq:defQ}
\end{equation}
where $\zetabar,\,\zeta$ are the quark fields at the boundary
$x_0=0$~\cite{definitions}. 
Finally the region $R$ is specified to be the space--time
volume between the hyperplanes $x_0 = t_1$ and $x_0 = t_2$.

On a lattice with finite spacing $a$, we then expect that
\begin{eqnarray}
   a^3\sum_{\bf x}
  \epsilon^{abc}
  \langle 
  [(\ar)^a_{0}(t_2,{\bf x})-(\ar)^a_{0}(t_1,{\bf x})] 
  \times \qquad\nonumber \\[-2ex]
  (\ar)^b_{k}(y) {\cal Q}_k^{c} 
  \rangle =  \qquad \nonumber \\[1ex]
 \;   2i
  \left\langle 
  (\vr)^c_{k}(y){\cal Q}_k^{c} 
  \right\rangle
  +\rmO(a^2), \;  t_1 < y_0 < t_2 \, ,
\label{eq:ward2}
\end{eqnarray}
is valid with correction terms of order $a^2$. 

Note that passing from eq.~(\ref{eq:ward1}) to eq.~(\ref{eq:ward2}) we
have specialized to $m=0$. The reason for
this is as follows. For finite quark mass, the second
term on the left hand side of eq.~(\ref{eq:ward1}) contains zero
separations $x - y$. Since it is not an integral over an on--shell
correlation function, it may have lattice artifacts of order $a$
and cannot be used to formulate an improvement condition. 

In terms of the bare unimproved correlation functions,
\begin{equation}
\kv(x_0) = -\frac{1}{9}\langle V_k^a(x){\cal Q}^a_k\rangle,
\end{equation}
\begin{equation}
\kt(x_0) = -\frac{1}{9}\langle T_{k0}^a(x){\cal Q}^a_k\rangle,
\end{equation}
as well as the bare improved correlation function,
\begin{equation}
\kaaI(x_0,y_0) = \frac{i}{18}\sum_{\bf x} \epsilon^{abc}
\langle (\aimpr)^a_0(x)(\aimpr)^b_k(y){\cal Q}^c_k\rangle,
\end{equation}
eq.~(\ref{eq:ward2}) may be rewritten, for $t_1 < x_0 < t_2$, as
\begin{eqnarray}
\zv[\kv(x_0)+a \cv \tilde{\partial}_0 \kt(x_0)] = \qquad \qquad \qquad \nonumber \\
\qquad \za^2[\kaaI(t_2,x_0) - \kaaI(t_1,x_0)] + \rmO(a^2) \, .
\label{eq:defcv}
\end{eqnarray}
At zero quark mass $m$, this equation can be solved for $\cv$.
However, due to a peculiarity of the quenched approximation, {\em i.e.} the
appearance of zero--modes in the quark propagator, it is not always
possible to simulate directly at zero quark mass~\cite{non_pert_act}. 
For the particular
boundary conditions and lattice size employed, here, this phenomenon is
relevant when $\beta=6/g_0^2 \leq 6.4$. In this region we have to
perform an extrapolation to zero quark mass. However, in the course of the
numerical simulations it turned out that $\cv$, implicitly 
defined by
eq.~(\ref{eq:defcv}) also for finite $m$, 
is a steep function of $m$ and could not be
reliably extrapolated to zero quark mass when we were not able to
simulate at very small masses. The reason for this strong dependence
is related to the fact that in the Ward identity~(\ref{eq:ward2})
the mass term was left out: as a consequence, eq.~(\ref{eq:defcv}) is
not  even valid in the continuum limit. 

\begin{figure}[t]
\vspace{-10pt}
\epsfig{file=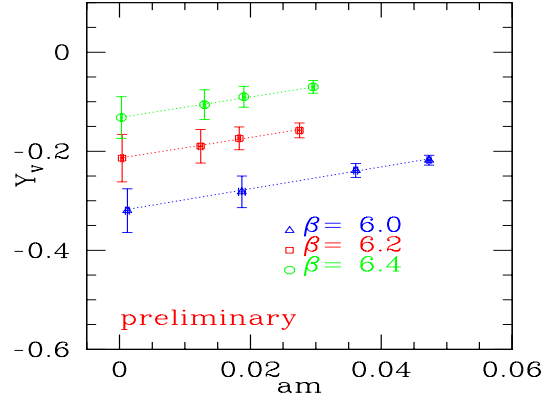, width=7cm, height=5.5cm}
\vspace{-20pt}
\caption{Dependence of $Y_{\rm V}$ on the quark mass.
}
\label{fig:mextra}
\end{figure}

It is therefore natural to repeat the
calculation, keeping the mass term. 
Eq.~(\ref{eq:ward2}) is then modified by an
additional term and has corrections 
of order $\rmO(am)$ instead of $\rmO(m)$. We  
solve the resulting equation formally for $\cv$ and denote the solution 
at finite mass by $Y_{\rm V}$, which coincides with $\cv$
as $m \rightarrow 0$.
As expected, this limit is now reached very smoothly (see 
figure~\ref{fig:mextra}). 

In a non--perturbative calculation, $\cv$ depends on the choices made
for the various kinematical parameters such as $T/L$ or the particular
Schr\"odinger Functional background field. Different choices lead to a
variation of $\cv$ itself that is O$(a)$. This is unavoidable and one
must make a specific choice; other choices would merely change
the size of the O$(a^2)$ effects after improvement. On the other hand,
one should search for values of the kinematical parameters such that
the O$(a^2)$ terms are small compared to the improvement term 
$a\tilde{\partial}_0 \kt(x_0)$. 
To achieve this, we 
first of all choose $t_1$ and
$t_2$ in such a way that the time separations between the various
fields in the correlation functions are as large as possible:
\begin{equation}
t_1 = T/4,\quad t_2 = 3T/4 \, .
\end{equation}
To fix the other parameters, we studied and tried to minimize
the cutoff effects in
eq.~(\ref{eq:defcv}) at tree level of perturbation theory (the
interested reader will find details in ref.~\cite{noi}).

With our final choices,
we observe reasonably
weak cutoff effects at tree level.
Apart from this perturbative study, we also performed a rough numerical
investigation at the largest values of the gauge coupling considering
different improvement conditions ({\em i.e.} different choices of
$L/a$ and other parameters), and observed the extracted values
for $\cv$ to be the same within
errors. 
The final definition of $\cv$ includes an average over the three
central time slices, $x_0$, which reduces statistical uncertainties.

\section{RESULTS}

\begin{figure}[htb]
\vspace{-5pt}
\epsfig{file=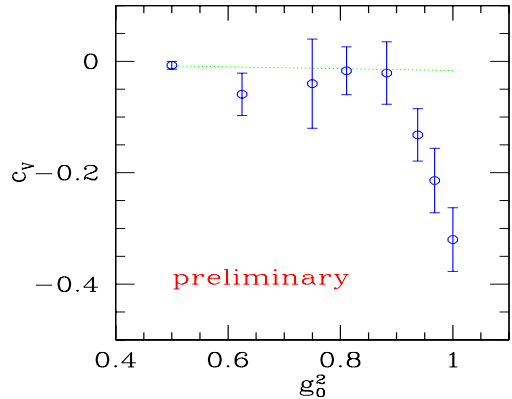, width=7cm, height=5.5cm}
\vspace{-20pt}
\caption{$c_V$ versus $g_0^2$. The dotted line is the 1--loop
  perturbative result.}
\label{fig:cv}
\end{figure}

Our preliminary results are summarized in figure~\ref{fig:cv}. Here we
show the non--perturbatively computed values of $\cv$ as a function of
the bare coupling $g_0^2$. The errors on $\cv$
receive a significant contribution from the (statistically independent) error
on the ratio $\za^2/\zv$.

In the region $\beta\leq 6.4$ we observe
sizeable values for $\cv$, far from 1--loop perturbation theory. 
In particular, at $g_0=1$ the
effect of the improvement term on the value of the vector meson decay
constant can be as large as 10\%.
When $g_0^2$ approaches 0, we obtain agreement with the
perturbative result. 

This work is part of the ALPHA collaboration research programme.
We thank DESY for allocating computer time for this project.

\begin{thebibliography}{9}

\bibitem{sym} K.~Symanzik, Nucl. Phys. B226 (1983) 187 and 205.

\bibitem{on_shell} M.~L\"uscher and P.~Weisz, Commun. Math. Phys. 97
  (1985) 59; Commun. Math. Phys. 98 (1985) 433

\bibitem{definitions} M.~L\"uscher, S.~Sint, R.~Sommer and P.~Weisz,
  Nucl. Phys. B478 (1996) 365. 

\bibitem{non_pert_act} M.~L\"uscher, S.~Sint, R.~Sommer, P.~Weisz and
  U.~Wolff, Nucl. Phys. B491 (1997) 323. 

\bibitem{one_loop} M.~L\"uscher and P.~Weisz, Nucl. Phys. B479 (1996)
  429. 

\bibitem{pert_cv} S.~Sint and P.~Weisz, hep--lat/9704001

\bibitem{non_pert_coeff} M.~L\"uscher, S.~Sint, R.~Sommer and H.~
  Wittig, Nucl. Phys. B491 (1997) 344. 

\bibitem{decay} M.~G\"ockeler {\em et al.}, hep--lat/9707021.

\bibitem{giulia} G. de Divitiis, talk at this conference.

\bibitem{schr} M.~L\"uscher, R.~Narayanan, P.~Weisz and U.~Wolff,
  Nucl. Phys. B384 (1992) 168; S.~Sint, Nucl. Phys. B421 (1994) 135,
  Nucl. Phys. B451 (1995) 416.

\bibitem{noi} M.~Guagnelli and R.~Sommer, work in progress.

\end{thebibliography}
\end{document}